\documentclass{moriond}

\usepackage{amsmath}
\usepackage{mathtools}
\usepackage{amsfonts}
\usepackage{amssymb}
\usepackage{bbold}

\usepackage{enumerate}

\usepackage{amssymb}
\usepackage{pifont}
\newcommand{\xmark}{\ding{55}}%

\def\be{\begin{equation}}
\def\ee{\end{equation}}
\def\bea{\begin{eqnarray}}
\def\eea{\end{eqnarray}}

\newcommand{\Photo}{}

\begin{document}
\vspace*{4cm}
\title{Leptoquark models for the $B$--physics anomalies}

\author{ Olcyr Sumensari }

\address{Laboratoire de Physique Th\'eorique (B\^at.~210),
CNRS and Univ. Paris-Sud, Universit\'e Paris-Saclay, 91405 Orsay cedex, France.\\[0.4em] 
Instituto de F\'isica, Universidade de S\~ao Paulo, \\
 C.P. 66.318, 05315-970 S\~ao Paulo, Brazil.}

\maketitle\abstracts{
The $B$-physics experiments at LHCb, BaBar and Belle hint towards deviations from Lepton Flavor Universality in both the tree-level and loop-induced $B$ meson semileptonic decays. We propose a leptoquark model with light right-handed neutrinos which can accommodate both $R_K^\mathrm{exp}<R_K^\mathrm{SM}$ and $R_{D^{(\ast)}}^\mathrm{exp}>R_{D^{(\ast)}}^\mathrm{SM}$. We discuss several of its predictions which can be tested in modern day experiments. We also comment on the recent finding at LHCb, namely $R_{K^\ast}^{\mathrm{exp}}<R_{K^\ast}^\mathrm{SM}$.}

\section{Introduction}

Even though no signal of New Physics appeared so far in the direct searches at the LHC, the $B$-physics experiments (BaBar, Belle and LHCb) hint at very intriguing deviations from lepton flavor universality (LFU). 
More specifically, the LHCb Collaboration measured the partial branching fractions of $B^+\to K^+\ell\ell$ in the bin $q^2 \in [1,6] \ \mathrm{GeV}^2$ and found~\cite{Aaij:2014ora}
\begin{align}\label{exp:RK}
R_K = \frac{ \mathcal{B}( B^+ \to K^+ \mu \mu)}{\mathcal{B}( B^+ \to K^+ e e)} = 0.745 \pm^{0.090}_{0.074} \pm 0.036 \,,
\end{align}
which lies $2.4 \sigma$ \underline{below} the Standard Model (SM) prediction, $R_K^{\rm SM} =1.00(1)$.~\cite{Bordone:2016gaq} Furthermore, another intriguing indication of LFU violation appeared in the tree-level processes, mediated by the charged currents,

\begin{align}
R_D = \left.\frac{ \mathcal{B}( B \to D \tau \nu)}{\mathcal{B}( B \to D \ell \nu)}\right|_{\ell \in \{e,\mu\}} \!\! = 0.41\pm0.05  \,,
\end{align}
which is obtained by combining several experimental values.~\cite{Lees:2012xj} This value appears to be $2.2 \sigma$ \underline{above} the SM prediction, $R_D^{\rm SM}=0.286\pm 0.012$, obtained by solely relying on the lattice QCD (LQCD) data for the form factors, recently computed in Ref.~\cite{MILC}. That result is corroborated by the experimentally established $R_{D^\ast}=0.310(15)(8)$, also confirmed by LHCb,~\cite{Aaij:2015yra} which appears to be $3.3 \sigma$ larger than the SM prediction, $R_{D^\ast}^{\mathrm{SM}}=0.252\pm 0.003$.~\cite{Fajfer:2012jt} Note, however, that the theoretical estimate of $R_{D^\ast}^{\rm SM}$ relies strongly on experimental information extracted from the differential distribution of $d\Gamma(B\to D^\ast \ell \nu)/dq^2$ (with $\ell=e,\mu$). The LQCD result for the full set of $B\to D^\ast$ form factors is still not available, and those are mandatory to consistently consider NP scenarios with couplings to both $\mu$ and $\tau$ (and not only to $\tau$-leptons), as suggested by current data.

Several models have been proposed to simultaneously accommodate $R_K$ and $R_{D^{(\ast)}}$, see Ref.~\cite{Becirevic:2016yqi} and references therein. While many authors considered effective scenarios, very few viable solutions to the puzzle of $B$--physics anomalies have been proposed. Among those, the models containing leptoquark (LQ) states are of particular interest as we will discuss in the following.

\section{Leptoquark models for $b\to s\ell\ell$}

Starting with the $R_K$ puzzle, the LQ states can be fully specified by their SM representation $(SU(3)_c,SU(2)_L)_{Y}$, where the hypercharge $Y$ is normalized by $Q=Y+T_3$. Among the LQ scenarios, the ones invoking vector LQs are not renormalizable and become problematic when computing the loop-induced processes, such as $\tau\to\mu\gamma$ and the $B_s\to \overline{B_s}$ mixing amplitude.~\cite{Fajfer:2015ycq} In Table~\ref{tab:lq-classification-RK}, we list the scalar LQ states that can modify $R_K$ through tree-level contributions to $b\to s \mu\mu$.~\cite{Becirevic:2016oho}
\begin{table}[htb!]
\renewcommand{\arraystretch}{1.3}
\centering
\begin{tabular}{|c|cccc|}
\hline 
\quad $(SU(3)_c,SU(2)_L)_{U(1)_Y}$  & BNC & Interaction & Eff.~Coefficients & $R_K/R_{K}^\mathrm{SM}$	 \\ \hline\hline
$(\bar{3},3)_{1/3}$	&   \xmark & $\overline{Q^C} i \tau_2 \boldsymbol{\tau}\cdot \boldsymbol{\Delta} L$	&$C_9=-C_{10}$	& $<1$ \\  
$(\bar{3},1)_{4/3}$	&   \xmark	& 	$\overline{d^C_R} \boldsymbol{\Delta}\ell_R$ &$(C_9)^\prime=(C_{10})^\prime$	&	$\approx1$ \\ 
$(3,2)_{7/6}$	&  \checkmark	& $\overline{Q}\boldsymbol{\Delta}\ell_R$	&	$C_9=C_{10}$ 	&	$>1$ \\  
$(3,2)_{1/6}$	&  \checkmark	&	$\overline{d_R}\widetilde{\boldsymbol{\Delta}}^\dagger L$	& $(C_9)^\prime=-(C_{10})^\prime$	& $<1$ 	\\  
\hline
\end{tabular}
\caption{\label{tab:lq-classification-RK}\small \sl List of LQ states which can modify $\mathcal{B}(B\to K \mu^+\mu^-)_{[1,6]~\mathrm{GeV}^2}$ at tree-level. The conservation of baryon number (BNC), the interaction term and the corresponding Wilson coefficients are also listed along with the prediction for $R_K$. Couplings to electrons are set to zero.}
\end{table}

\noindent From this Table we see that only the states $(3,2)_{1/6}$ and $(\bar{3},3)_{1/3}$ can consistently accommodate $R_K^\mathrm{exp}<R_K^\mathrm{SM}$ at tree-level. Notice, however, that the latter state violates baryon number via the dangerous diquark couplings, which can induce the proton decay at tree-level.~\cite{Dorsner:2016wpm} In the following we discuss how the scenario $(3,2)_{1/6}$, originally proposed in Ref.~\cite{Becirevic:2015asa}, can be consistently extended to accommodate $R_K$ and $R_{D}$ without contradictions with other flavor physics constraints.

\section{A leptoquark model to explain $R_K$ and $R_D$}

In Ref.~\cite{Becirevic:2016yqi}, it was pointed-out that the inclusion of light right-handed (RH) neutrinos to the model $(3,2)_{1/6}$ induces new contributions to charged current processes. The Lagrangian of the LQ model then becomes 
\begin{align}
\label{eq:slq2-nuR}
\begin{split}
\mathcal{L}_{\Delta^{(1/6)}} &= Y_L^{ij} \overline{d}_{Ri} {\widetilde{\boldsymbol{\Delta}}}^{\dagger} L_j+Y_R^{ij} \overline{Q}_{i} \boldsymbol{\Delta}^{} \nu_{R,j}+\mathrm{h.c.},
\end{split}
\end{align}
\noindent where $i,j$ stand for flavor indices and $y_{L,R}$ are two generic Yukawa matrices. The LQ doublet is denoted by $\boldsymbol{\Delta}$ and we define the left-handed doublets as $Q_i=[(V^\dagger u_L)_i\; d_{Li}]^T$ and $L_j=[(U\nu_L)_i\;\ell_{Li}]^T$, where $V$ and $U$ are the Cabibbo-Kobayashi-Maskawa (CKM) and Pontecorvo-Maki-Nakagawa-Sakata (PMNS) matrices, respectively. We reiterate that the novelty of this model is the introduction of the second term in Eq.~\eqref{eq:slq2-nuR} which induces the LQ interaction with up-type quarks.

In the following we assume that the two LQ are mass-degenerate, $m_{\Delta} \approx 1~\mathrm{TeV}$, and that the RH neutrinos are massless in comparison with the hadronic scale. The transitions $b\to s\ell\ell$ and $b\to c\ell \nu$ can then be described by a low-energy effective theory with interaction terms
\begin{align}\label{eq:Lbsmm}
&\mathcal{L}^{d_k\to d_i\ell\ell}_{\mathrm{eff}} = - \frac{Y_L^{ij}  Y_L^{\ast kl}  }{2 m_\Delta^2}\ \overline d_i\gamma_\mu P_R d_k \ \overline \ell_l \gamma^\mu P_L\ell_j +\mathrm{ h.c.} \,,
\end{align}
\noindent and
\begin{align}\label{eq:Leff2}
\mathcal{L}^{d\to u\ell\overline \nu}_{\mathrm{eff}} = & \frac{(V\cdot Y_R)^{ij}  Y_L^{\ast kl}  }{2 m_\Delta^2}   \left[ 
\overline u_i P_R d_k \ \overline \ell_l P_R\nu_j +\frac{1}{4} \overline u_i \sigma_{\mu\nu}P_Rd_k\  \overline \ell_l \sigma^{\mu\nu} P_R\nu_j \right] + \mathrm{ h.c.}\,,
\end{align}
which will be used in the phenomenogical discussions below. Notice that the contributions to the charged processes (and to $b\to c\ell\nu$ in particular) depend on the existence of RH neutrinos.

\section{Constraints and predictions}
\label{sec:constraints}

For simplicity, the couplings to the first generation are set to zero to avoid the potential problems with the atomic parity violation experiments,~\cite{Dorsner:2016wpm} as well as the experimental limits on $\mathcal{B}(K\to\pi\nu \nu)$ and $\mathcal{B}(B_s\to \mu e)$.~\cite{Becirevic:2016yqi} The couplings are varied within the perturbativity limit, $|(y_L)_{ij}|\leq 4\pi$, and are confronted with several constraints of which the most relevant ones are: (i) the experimentally established $\mathcal{B}(B_s\to\mu\mu)$, and $\mathcal{B}(B\to K\mu\mu)$ in the $[15,19]~\mathrm{GeV}^2$ bin, (ii) $B_s - \overline{B}_s$ mixing, (iii) bounds on the lepton flavor violating $\tau$ decays, such as $\mathcal{B}(\tau\to\mu\phi)$ and $\mathcal{B}(\tau\to\mu\gamma)$, (iv) (semi--)leptonic meson decays, (v) the ratio $R_D^{\mu/e}=\mathcal{B}(B\to D\mu\nu)/\mathcal{B}(B\to De\nu)$, and (vi) limits on $\mathcal{B}(B\to K\nu\nu)$, cf.~Ref.~\cite{Becirevic:2016yqi} for details. 

\begin{figure*}[h]
\begin{center}
\begin{tabular}{ccc}
\includegraphics[width=0.31\textwidth]{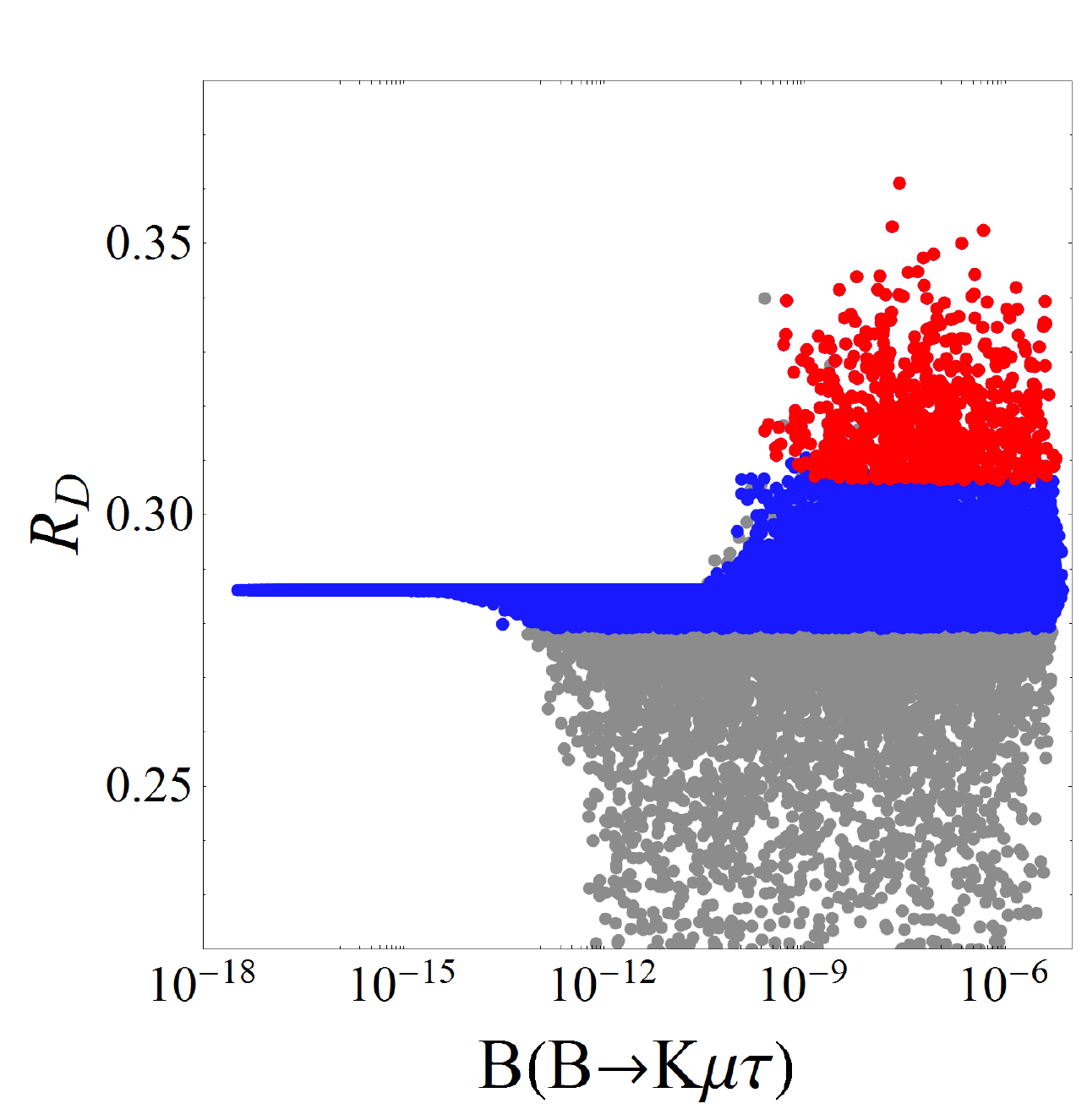}
\includegraphics[width=0.31\textwidth]{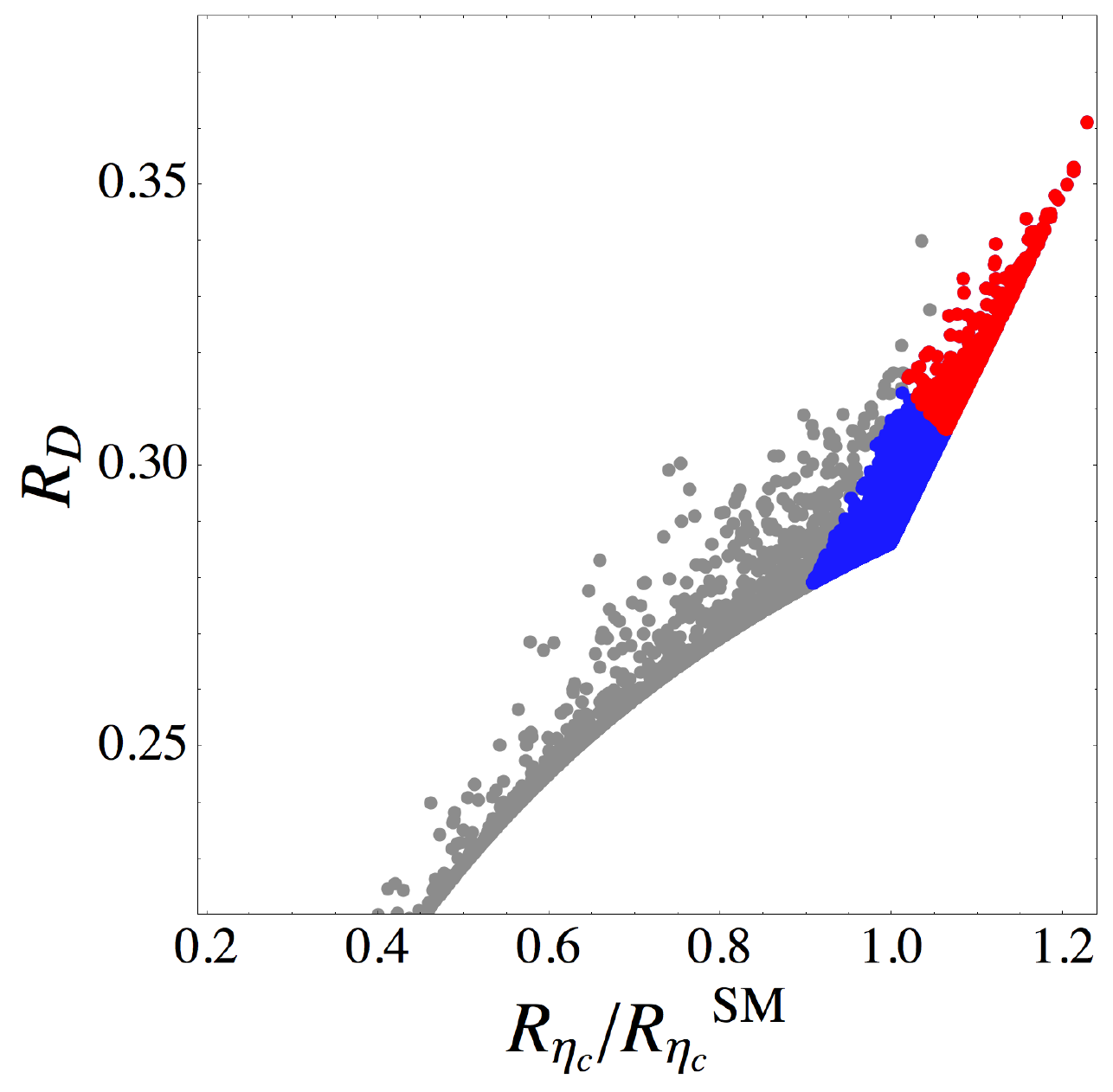}
\includegraphics[width=0.31\textwidth]{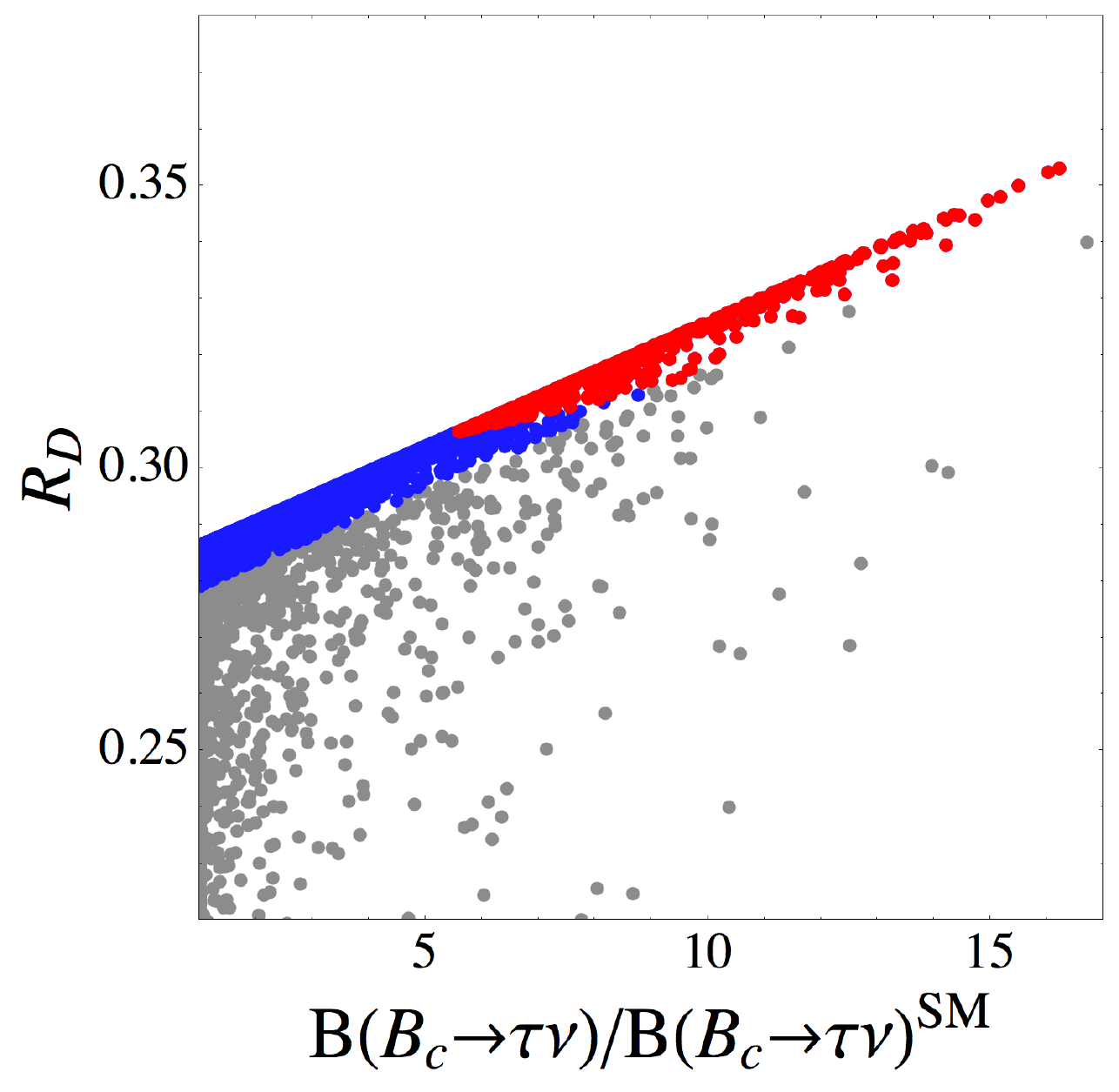}
\end{tabular}
\end{center}
\caption{The blue points are obtained by subjecting the Yukawa couplings of our model to the constraints discussed in Sec.~\ref{sec:constraints}, and the red ones are selected from the blue ones after requiring the compatibility with 
$R_D^{\rm exp}$ to $2\sigma$. We plot our predictions for three selected quantities: $\mathcal{B}(B\to K\mu\tau)$, the ratio between $R_{\eta_c}=\mathcal{B}(B_c\to \eta_c\tau\bar \nu)/\mathcal{B}(B_c\to \eta_c l\bar \nu)$ predicted by our model and its SM value, and a similar ratio of $\mathcal{B}(B_c\to \tau\bar \nu)$. \label{fig:2}}
\end{figure*}

After applying the constraints described above, we find that this model can not only predict $R_K=0.88(8)$, compatible with the experimental finding, but also accommodate the excess in $R_D$ at the $1\sigma$ level. In other terms, our model can satisfactorily explain the anomalies $R_K$ and $R_D$. Notice that we only focus on $R_D$ because all the needed form factors have been computed on the lattice.~\cite{Becirevic:2016yqi} We cannot provide the accurate statement concerning $R_{D^\ast}$ because the full set of $B\to D^\ast$ form-factors is not available from LQCD simulations. We can only make a qualitative observation that $R_{D^\ast}>R_{D^\ast}^\mathrm{SM}$ in our model. Our main predictions are shown in Fig.~\ref{fig:2} and summarized below:~\cite{Becirevic:2016yqi}

\vspace*{-0.4cm}

\begin{itemize}
	\item[$\bullet$] We computed the LFV decay $\mathcal{B}(B\to K\mu\tau)$, which is found to be 
	\begin{equation}
		2.1\times 10^{-10} \leq \mathcal{B}(B\to K\mu\tau) \leq 6.7\times 10^{-6},
	\end{equation}		
	\noindent also shown in Fig.~\ref{fig:2}. The predictions for the other LFV modes can be inferred bound given above via the relations $\mathcal{B}(B_s\to\mu\tau)\approx 0.9 \times \mathcal{B}(B\to K\mu\tau)$ and $\mathcal{B}(B\to K^\ast\mu\tau)\approx 1.8 \times \mathcal{B}(B\to K\mu\tau)$ derived in Ref.~\cite{Becirevic:2016zri}.
	
	\item[$\bullet$] A distinctive prediction of the model is that the ratio $R_{\eta_c}=\mathcal{B}(B_c\to\eta_c\tau\nu)/\mathcal{B}(B_c\to\eta_c\ell\nu)$ and the leptonic decay mode $\mathcal{B}(B_c\to\tau\nu)$ can be considerably larger than the SM predictions. We found that
	\begin{equation}
		1.02 \leq R_{\eta_c}/R_{\eta_c}^\mathrm{SM} \leq 1.21,\qquad \text{and} \qquad 5.5 \leq \mathcal{B}(B_c\to \tau \nu) \leq 16.1,
	\end{equation}
\noindent as shown in Fig.~\ref{fig:2}, which offer an alternative experimental test of the validity of our model.
\end{itemize}

\vspace*{-0.6cm}

\paragraph{Can the $R_{K^\ast}$ hints be explained by scalar leptoquarks?}
\mbox{}\\[0.5em]
\noindent A preliminary result for $R_{K^\ast}$ has been presented by LHCb which indicates another deviation from LFU. The ratio $R_{K^\ast}=\mathcal{B}(B\to K^\ast \mu\mu)/\mathcal{B}(B\to K^\ast ee)$ in two different $q^2$ bins appears to be $2.2-2.4~\sigma$ below the SM prediction.~\cite{Bifani} If confirmed, this result would exclude the model discussed above, since it predicts $R_{K^\ast}$ to be slightly larger than $R_{K^\ast}^\mathrm{SM}$. The only LQ state that can explain both $R_K^\mathrm{exp}<R_K^\mathrm{SM}$ and $R_{K^\ast}^\mathrm{exp}<R_{K^\ast}^\mathrm{SM}$ at tree-level is the $SU(2)_L$ triplet $(\bar{3},3)_{1/3}$.~\cite{Crivellin:2017zlb} Nonetheless, as discussed above, an additional symmetry is needed to forbid dangerous diquark couplings from destabilizing the proton.~\cite{Dorsner:2016wpm} Another possibility recently proposed is to consider the doublet LQ $(3,2)_{7/6}$ amended with a symmetry to forbid the tree-level contribution to $b\to s\ell\ell$.~\cite{Becirevic:2017jtw} This latter scenario generates the Wilson coefficients $C_9=-C_{10}<0$ through loops and it has the great advantage of not disturbing the proton stability.

\section{Conclusions}

In this proceeding we discussed a LQ model which can explain the LFU anomalies in both charged and neutral $B$-meson decays, namely $R_K^\mathrm{exp}<R_K^\mathrm{SM}$ and $R_{D^{(\ast)}}^\mathrm{exp}>R_{D^{(\ast)}}^\mathrm{SM}$. Our model offer several predictions which can be tested in the near future: (i) branching ratios for the exclusive $b\to s\mu\tau$ modes can be as large as $\mathcal{O}(10^{-6})$, being also bounded from below; (ii) the LFUV effects in $R_{\eta_c}=\mathcal{B}(B_c\to\eta_c\tau\nu)/\mathcal{B}(B_c\to\eta_c\ell\nu)$ can be larger than predicted in the SM, and (iii) $\mathcal{B}(B_c\to\tau\nu)$ is predicted to be enhanced by a factor of $5\div 16$ with respect to the SM value. Furthermore, we devise a scalar LQ model which can explain $R_{K^{(\ast)}}^\mathrm{exp}<R_{K^{(\ast)}}^\mathrm{SM}$ through loop effects.

\section*{Acknowledgments}

This project has received funding from the European Union's Horizon 2020 research and innovation program under the Marie Sklodowska-Curie grant agreements No.~690575 and No.~674896.  


\end{document}